\documentclass[aps,prd,twocolumn,superscriptaddress,showpacs,nofootinbib]{revtex4-1}

\usepackage{epsfig}
\usepackage{graphicx}
\usepackage{dcolumn}
\usepackage{bm}
\usepackage{multirow}
\usepackage{array}
\usepackage{slashed}
\usepackage{amsmath}
\usepackage[titletoc]{appendix}
\usepackage{color}
\usepackage{hyperref}
\usepackage[normalem]{ulem}
\usepackage{soul}

\begin{document}

\title{Signature of Collapsars as Sources for High-energy Neutrinos and $r$-process Nuclei} 

\author{Gang Guo}
\email{guogang@cug.edu.cn} 
\affiliation{School of Mathematics and Physics, China University of Geosciences, Wuhan 430074, China}
\author{Yong-Zhong Qian}
\email{qian@physics.umn.edu} 
\affiliation{School of Physics and Astronomy, University of Minnesota, Minneapolis, MN 55455, USA}
\author{Meng-Ru Wu}
\email{mwu@gate.sinica.edu.tw}
\affiliation{Institute of Physics, Academia Sinica, Taipei, 11529, Taiwan}
\affiliation{Institute of Astronomy and Astrophysics, Academia Sinica, Taipei, 10617, Taiwan}

\date{\today}

\begin{abstract}
If collapsars are sources for both high-energy (HE) neutrinos and $r$-process nuclei,
the profuse low-energy antineutrinos from $\beta$-decay of the newly-synthesized nuclei 
can annihilate the HE neutrinos. Considering HE neutrinos produced at internal shocks induced 
by intermittent mildly-magnetized jets, we show that such annihilation suppresses the overall
HE neutrino spectrum at $\gtrsim 300$~TeV and produces a corresponding flavor composition of 
$(F_{\nu_e+\bar\nu_e}: F_{\nu_\mu+\bar\nu_\mu}: F_{\nu_\tau+\bar\nu_\tau})_\star \approx (1 : 10 : 1)$
at source. We find that the emergent HE neutrino flux can well fit the diffuse flux 
observed at IceCube if contributions from all similar 
sources are taken into account. Our results highlight the unique role of HE neutrinos in
supporting collapsars as sources for $r$-process nuclei, and can be tested by detection of HE neutrinos 
from individual sources and accurate measurement 
of the diffuse HE neutrino flux spectrum and flavor composition.
\end{abstract}


\maketitle

\section{Introduction}
Collapsars produced by the collapse of massive stars into black holes
have long been considered a leading candidate for powering Type Ibc supernovae and long gamma-ray bursts (GRBs)
of both the bright and low-luminosity (LLGRBs) varieties \cite{1993ApJ...405..273W,MacFadyen:1999xx,Woosley:2006fn,Campana:2006xx,Hjorth:2012xx,Kumar:2014upa}.
Shocks associated with the propagation of the collapsar jet through the stellar envelope and/or 
the black-hole accretion disk wind were also proposed as candidate sites for producing $\sim$~TeV--PeV
high energy (HE) neutrinos 
\cite{Meszaros2001,Razzaque:2003uv,Razzaque2004,Ando:2005xi,Razzaque2005,Horiuchi:2007xi,Enberg:2008jm,Bartos:2012sg,Murase2013,Fraija2013,Xiao2014,Bhattacharya2014,Varela2015,Tamborra2015,Senno2015,Senno:2017vtd,Denton:2017jwk,Denton:2018tdj,He:2018lwb,Chang:2022hqj,Guarini:2022hry}.
These neutrinos may contribute significantly to the diffuse flux detected by IceCube, whose astrophysical origin 
remains unknown despite recent reports of potential association of several events with a blazar~\cite{Aartsen2018,Aartsen:2018cha}, three tidal disruption events~\cite{Stein:2020xhk,Reusch:2021ztx,vanVelzen:2021zsm}, and an active galaxy~\cite{IceCube:2022der}.

The ground-breaking multi-messenger observations of GW170817 linked binary neutron star mergers (BNSMs) to
both short GRBs and the production of heavy elements beyond iron via the rapid neutron-capture process ($r$-process)~\cite{LIGOScientific:2017vwq,LIGOScientific:2017ync,Kasen:2017sxr,Drout:2017ijr,Cowperthwaite:2017dyu,Villar:2017wcc,Shibata:2017xdx,Metzger:2019zeh,Watson:2019xjv}. The astrophysical environments of BNSMs and collapsars are rather similar in that both host
accretion disks and the associated jets, thereby producing GRBs and potentially HE neutrinos.
Interestingly, Ref.~\cite{Siegel:2018zxq} showed that the physical conditions of disk outflows in collapsars 
resemble those of outflows in BNSMs, and proposed that collapsars are a likely or even the dominant site for 
producing $r$-process nuclei. This proposal remains hotly debated 
and requires further theoretical and observational efforts to clarify~\cite{Siegel:2019mlp,Macias:2019oxw,Bartos:2019twj,Miller:2019mfl,Fujibayashi:2020jfr,Brauer:2020hty,Fraser:2022xx,Barnes:2022dxv,Just:2022fbf,Fujibayashi:2022xx,Anand:2023ujd}.

\begin{figure}[tbp]
\includegraphics[width=0.9\columnwidth]{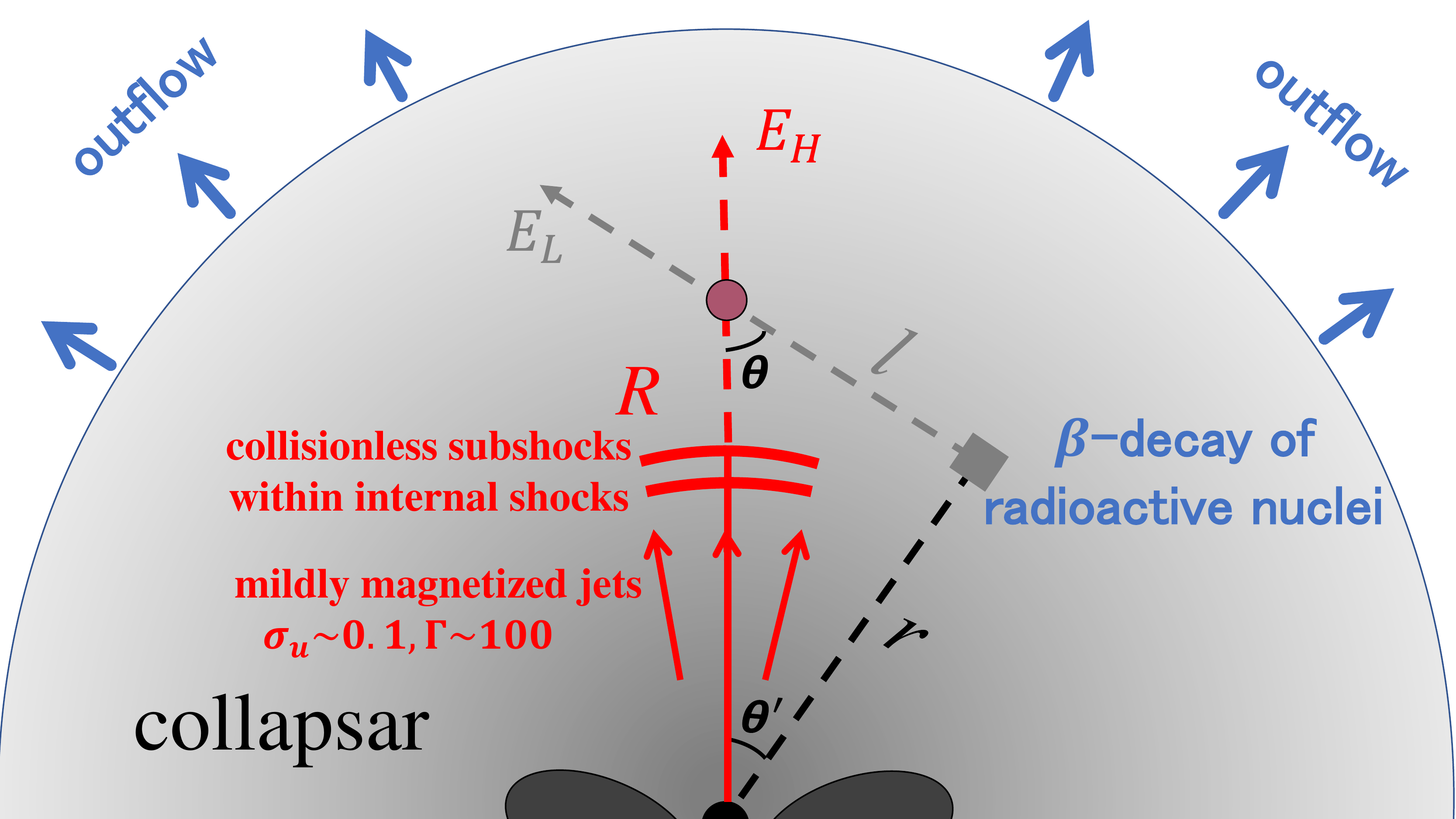}
\caption{Sketch of production of HE neutrinos (with energy $E_H$) and their annihilation with LE antineutrinos (with energy $E_L$) from
$\beta$-decay of $r$-process nuclei in a collapsar.}
\label{fig:sketch} 
\end{figure}

In this work, we investigate for the first time a novel connection between HE neutrinos and $r$-process nucleosynthesis in collapsars. 
Unstable neutron-rich heavy nuclei are produced by rapid neutron capture during the $r$-process. Their $\beta$ decay produces $\bar\nu_e$ 
with energy $E_L\sim 4$~MeV on a timescale of $\sim 1$~s \emph{everywhere} inside the outflow that expands with a typical velocity 
$v_{\rm ej}\sim (0.05$--$0.3)c$ (see Fig.~\ref{fig:sketch}). Following flavor oscillations, these low energy (LE) antineutrinos can 
annihilate HE neutrinos of the corresponding flavor produced by shocks at radius $R$. 
Efficient annihilation via the $Z$-resonance requires $s=2 E_H E_L (1-\cos\theta)\sim m_Z^2$,
where $\theta\sim\mathcal{O}(1)$ is the intersection angle and $m_Z$ is the $Z$ mass. Therefore, the HE neutrino flux at $E_H\gtrsim\mathcal{O}(100$--1000)~TeV
is expected to be affected. As we will show, such annihilation can leave clear imprints on the energy spectrum and flavor composition of 
the emergent HE neutrinos. 

\begin{table*}[htbp] 
\centering 
\caption{Proper values of magnetic field $B_{d(u)}$, proton number density $n_{p,d(u)}$, and photon temperature $T_{\gamma,d(u)}$ in the shocked (downstream) and unshocked (upstream) jets.}
\renewcommand{\arraystretch}{1.5}
\begin{tabular}{lccc}
\hline
\hline
& $B_{d(u),8}$ & $n_{p,d(u),19}$ & $T_{\gamma,d(u),{\rm keV}}$  \\     
\hline
downstream & $8.2(\epsilon_{B,d}L_{{\rm iso},52})^{1/2}/(R_{{\rm is},10}\Gamma_2)$ & $1.8 L_{\rm iso, 52}(1-\epsilon_{e,d}-\epsilon_{B,d})R_{\rm is,10}^{-2}\Gamma_2^{-2}$ &
$3.7 (\epsilon_{e,d} L_{\rm iso, 52})^{1/4}R_{\rm is,10}^{-1/2}\Gamma_2^{-1/2}$
\\
upstream & $B_{d,8}/\xi$ &  $n_{p,d,19}/\xi$  & $T_{\gamma,d,{\rm keV}}(\epsilon_{e,u}/\epsilon_{e,d})^{1/4}(\Gamma/\Gamma_{r})^{1/2}$ \\
\hline
\end{tabular}
\label{tab:cond} 
\end{table*} 

\section{Antineutrinos from $\beta$-decay}
The $\beta$-decay of the newly-synthesized $r$-process nuclei not only produces a profuse
$\bar\nu_e$ flux, but also provides a dominant source of energy to heat the associated ejecta \cite{Metzger:2010}. We estimate the
$\bar\nu_e$ emission using the power generated by $\beta$-decay. In our benchmark study, we simply assume that the ejecta expands with a constant velocity $v_{\rm ej}$, forming a steady spherical ``wind'' with a constant mass outflow rate $\dot M$ (see Appendix~\ref{app:geometry}).
As the ejecta expands, the power generated by $\beta$-decay per unit mass stays almost constant for a period of $\sim T_r$
when the $r$-process produces neutron-rich nuclei far from stability. After neutron capture ceases, that power
approximately follows a power-law decline.
We take $\dot\epsilon_{\bar\nu_e,0}\eta(t)$ as the part of the power per unit mass that is carried away by $\bar\nu_e$, where $\dot\epsilon_{\bar\nu_e,0}$ sets the magnitude and $\eta(t)$ characterizes the time evolution with $\eta(0)\approx 1$. From numerical calculations, $\eta(t)$ can be fitted as \cite{Rosswog:2013kqa} 
\begin{align}
\eta(t) \approx \left[ \frac{1}{2} - \frac{1}{\pi} \arctan\left(\frac{ t-T_{r}}{0.11~\rm s}\right)\right]^{1.3}.
\end{align}
Ignoring nuclear shell effects and Coulomb correction, we simply take the $\bar\nu_e$ spectrum to be $\propto E^2(Q-E)^2$ with $Q$ being the $Q$-value. The energy-differential emission rate of $\bar\nu_e$ per unit mass (i.e., emissivity) 
at radius $r=v_{\rm ej}t$ can be estimated as
\begin{align}
 j_{\bar\nu_e}(E,r) &=\frac{\dot\epsilon_{\bar\nu_e,0}\eta(r/v_{\rm ej})}{\langle E_L \rangle}
 \left[\frac{15(2\langle E_L \rangle-E)^2 E^2}{16\langle E_L \rangle^5}\right],
\end{align}
where the term in the brackets is the approximate normalized $\bar\nu_e$ spectrum with an average energy of
$\langle E_L \rangle=Q/2=4$~MeV.
Note that both $\dot\epsilon_{\bar\nu_e,0}$ and $T_r$ depend on the electron fraction $Y_e$ of the ejecta \cite{Rosswog:2013kqa}. 
Assuming a typical value of $Y_e=0.2$ for collapsar outflows and guided by the power used in Ref.~\cite{Wu:2018mvg}, we take 
$\dot\epsilon_{\bar\nu_e,0}=5 \times 10^{18}$~erg/g/s and $T_r=0.4$~s. 

The $\bar\nu_e$ emission is approximately isotropic.
Referring to Fig.~\ref{fig:sketch}, we estimate the $\bar\nu_e$ 
intensity (in units of cm$^{-2}$~s$^{-1}$~MeV$^{-1}$~sr$^{-1}$) at radius $R$ and angle $\theta$ by integrating the emissivity 
along the path length $l$:
\begin{align}
I_{\bar\nu_e}(E,R,\theta)&=\int\frac{\rho(r)j_{\bar\nu_e}(E,r)}{4\pi}dl\nonumber\\
&=\frac{R\dot M}{16\pi^2v_{\rm ej}}\int\frac{j_{\bar\nu_e}(E,r)\sin\theta}{r^2\sin^2(\theta+\theta')}d\theta',
\label{eq:Inu}
\end{align}
where $\rho(r)=\dot M/(4\pi r^2v_{\rm ej})$ is the mass density of the ejecta at radius $r$, and $r$ is
a function of $R$, $\theta$, and $\theta'$.
Unless noted otherwise, we take $\dot M=0.02\,M_\odot/{\rm s}$ \cite{Hayakawa:2018uxm,Fujibayashi:2022xx} and $v_{\rm ej}=0.05c$ \cite{MacFadyen:1999xx,Miller:2019mfl}.
The corresponding total number density of LE $\bar\nu_e$ at $R\sim 10^{10}$~cm 
is $\sim 10^{23}~{\rm cm}^{-3}$. 

Following production, the LE $\bar\nu_e$ undergo flavor oscillations prior to interacting with the HE neutrinos. The oscillation scenario depends on the matter densities at the $\bar\nu_e$ emission site and the interaction site. We consider three sets of probabilities $P_{\bar\nu_\alpha}$ for the initial $\bar\nu_e$ to be a $\bar\nu_\alpha$ at the interaction site: $(P_{\bar\nu_e}, P_{\bar\nu_\mu}, P_{\bar\nu_\tau})\approx (0.55, 0.18, 0.27)$ (pure vacuum oscillations), $(0.675, 0.095, 0.23)$ (NH), and $(0.022, 0.545, 0.433)$ (IH). The latter two sets assume adiabatic flavor evolution following emission at high densities for normal (NH) and inverted (IH) neutrino mass hierarchy, respectively (see Appendix~\ref{app:oscil}).

\begin{figure*}[htbp]
\includegraphics[width=1.8\columnwidth]{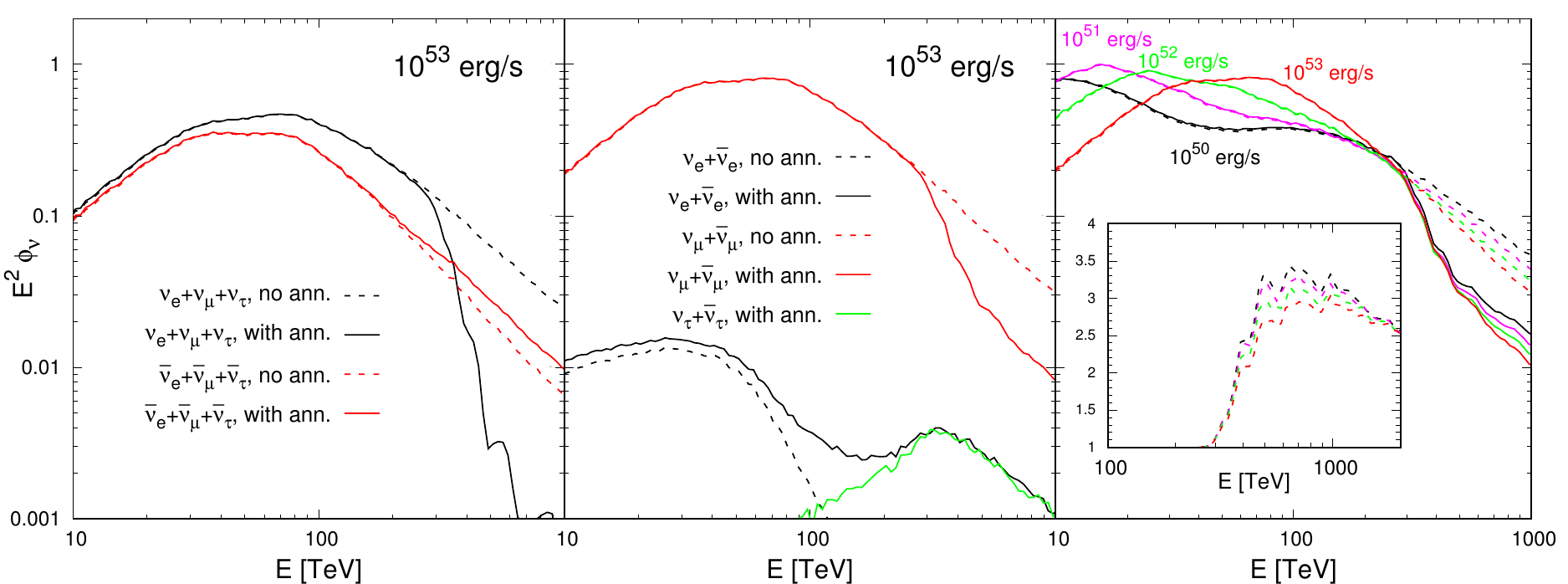}
\caption{HE neutrino spectra including annihilation with LE antineutrinos (solid curves) or not (dashed curves). Vacuum oscillations of LE antineutrinos are assumed. The benchmark case with $L_{\rm iso}=10^{53}$~erg/s is shown in the left ($\nu_e+\nu_\mu+\nu_\tau$ and $\bar\nu_e+\bar\nu_\mu+\bar\nu_\tau$) and middle ($\nu_\alpha + \bar\nu_{\alpha}$ with $\alpha=e, \mu, \tau$) panels. Cases of 
$L_{\rm iso}=10^{50}$--$10^{53}$~erg/s are shown in the right panel (all flavors). The inset shows the ratios of all-flavor spectra without annihilation to those with annihilation.}
\label{fig:nu_flavor}    
\end{figure*}

\section{High-energy neutrino production}
Relativistic jets in collapsars induce shocks that can accelerate electrons and protons. 
The accelerated electrons can produce $\gamma$-rays or X-rays through synchrotron radiation or inverse Compton scattering, 
thereby making bright GRBs, while the HE protons can collide with photons or stellar matter to make $\pi^\pm$ and $K^\pm$, thereby 
producing HE neutrinos through meson decays. Shocks emerging at different sites during jet propagation lead to different 
scenarios for HE neutrino production 
\cite{Waxman1997,Waxman:1999ai,Li2002,Guetta:2003wi,Murase:2006dr,Murase:2006mm,Murase:2007yt,Hummer:2011ms,Zhang:2012qy,Liu2013,Bustamante2015,Nir:2015teo,Biehl:2017zlw,Guo:2019ljp,Kimura:2022zyg}. To demonstrate the effects of LE antineutrinos from $\beta$-decay, we focus on HE neutrinos 
produced during jet propagation deep inside the progenitors of collapsars. 
Specifically, we consider proton acceleration at internal shocks before jet collimation \cite{Meszaros2001,Murase2013}. 

Ref.~\cite{Murase2013} emphasized that shocks inside stars are likely mediated by radiation and in contrast to
collisionless shocks \cite{Levinson:2007rj,Katz:2010}, may not support efficient particle acceleration and HE neutrino production.
It was shown later, however, that if the jets are mildly magnetized, a strong collisionless subshock may occur within the 
radiation-mediated shock (RMS) \cite{Beloborodov:2016jmz,Lundman:2018mad} to allow particle acceleration at, e.g., mildly relativistic 
internal shocks \cite{Crumley:2018kvf}. Indeed, using relativistic magnetohydrodynamic simulations of magnetized jets from BNSMs, 
Ref.~\cite{Gottlieb:2021pzr} found that subshocks can emerge at small radii and produce HE neutrinos at internal shocks, 
collimation shocks, and shock breakout.          

Interestingly, a mild magnetization is also essential to avoid heavy baryon loading of the jets from mixing with the cocoon material \cite{Gottlieb:2020mmk,Gottlieb:2021avb,Gottlieb:2021pzr}. With negligible jet-cocoon mixing, the jet Lorentz factor could grow almost linearly with radius due to adiabatic expansion (see, e.g., \cite{Piran2004,Meszaros2006,Kumar2014}) and reach $\gtrsim100$ deep inside the star \cite{Mizuta:2008ch,Gottlieb:2021avb}.
Internal shocks caused by collisions of slow and rapid jets can occur at $R_{\rm is} \approx \Gamma_s^2 c \delta t = 3 \times 10^{9} \Gamma^2_{s, 1} \delta t_{-3}$ cm, where $\Gamma_s$ is the typical Lorentz factor of the slow jets and $\delta t$ is the variability timescale. Here and below, $A_x=A/10^x$, where $A$, if dimensional, is in cgs units. We assume that the typical Lorentz factor $\Gamma_r$ of the unshocked rapid jets is twice the Lorentz factor $\Gamma$ of the shocked jet. The relative Lorentz factor between the two is $\Gamma_{\rm rel} \approx \Gamma_r/(2\Gamma) + \Gamma/(2\Gamma_r)=1.25$, consistent with the mildly relativistic shock. 

In our scenario, a strong collisionless subshock forms within the internal RMS to facilitate particle acceleration. This scenario requires $\sigma_u \equiv B_u^2/(4\pi\rho_u c^2) \gtrsim 0.01$ \cite{Beloborodov:2016jmz,Lundman:2018mad}, where $B_u$ and $\rho_u$ are the proper magnetic field and mass density, respectively, of the unshocked upstream region. Due to shock compression, the magnetic field
is amplified to $B_d=\xi B_u$ in the shocked downstream region with the compression ratio $\xi\sim 5$ for $\Gamma_{\rm rel}\approx 1.25$ \cite{Beloborodov:2016jmz}. Because HE neutrino flux would be strongly suppressed by synchrotron cooling of $\pi^\pm$ and $K^\pm$ in the enhanced magnetic field $B_d$, we focus on HE neutrino production via $pp$ and $p\gamma$ reactions in the upstream region (i.e., the unshocked jet), which was barely discussed in previous literature. For radiation-dominated jets at launching, the initial jet temperature is $\sim 1.3 L^{1/4}_{\rm iso, 52}$ MeV at $r=10^7$ cm \cite{Kumar:2014upa}. The jets expand adiabatically and the temperature decreases as $r^{-1}$. At $r=10^9$--$10^{10}$ cm, the jet temperature remains a few keV, corresponding to a thermal energy fraction $\epsilon_{e,u} \gtrsim 0.1$.
Proper values of magnetic field, proton number density, and photon temperature in the shocked and unshocked jets are given in Table~\ref{tab:cond}.

Protons can be accelerated to high energies by crossing the shock fronts repeatedly. During this process, protons have significant probabilities to escape from both the upstream and downstream regions, resulting in a power-law spectrum \cite{Bell1978}. Recent simulations of mildly relativistic shocks showed that the accelerated protons in the upstream region are much more abundant than those in the downstream region (see Figs. 7 and 8 of \cite{Crumley:2018kvf}).
To obtain a conservative estimate of the maximal proton energy $E_{p,{\rm max}}$, we compare 
the timescales of shock acceleration and different cooling processes for protons in the shocked jet 
(see Appendix~\ref{app:timescale}).
For all the parameter sets adopted in this work, the $E_{p,{\rm max}}$ computed in the rest frame of the shocked jet is always sufficient to make PeV neutrinos ($E_\nu\sim 0.05\Gamma E_p$ from $pp$ and $p\gamma$ reactions).

We take the (unnormalized) spectrum of the accelerated protons to be $\phi_p(E_p) = E_p^{-2}\exp(-E_p/E_{p, \rm max})$, and use PYTHIA 8.3 \cite{Bierlich:2022pfr} and SOPHIA \cite{Mucke:1999yb} for $pp$ and $p\gamma$ reactions, respectively, to obtain the $\pi^\pm$ and $K^\pm$ yields at different centre-of-mass energies.
Because the shocked jet is only mildly relativistic relative to the rapid jet ($\Gamma_{\rm rel}=1.25$), we ignore the Lorentz transformation for $E_p$ and $\phi_p(E_p)$ in our calculations.
We include cooling of $\pi^\pm$ and $K^\pm$ due to synchrotron radiation, inverse Compton scattering, $e^\pm$ pair production on photons ($\pi^\pm/K^\pm + \gamma \to \pi^\pm/K^\pm + e^+ + e^-$), hadronic scattering, and adiabatic expansion (see Appendix~\ref{app:timescale}).

In our benchmark study, we take $R_{\rm is}=3\times 10^{9}$ cm, $L_{\rm iso}=10^{53}$ erg/s, $\Gamma_r=2\Gamma\approx 90L^{0.18}_{\rm iso, 49}\approx 472$ \cite{Yi:2016hlp}, $\sigma_d =\xi\sigma_u \approx 2\epsilon_{B,d}=0.25$, $\xi=5$, $\epsilon_{e,d}=0.5$, and $\epsilon_{e,u}=0.3$ (see Table~\ref{tab:cond}). The corresponding $E_{p,{\rm max}}$ is $\approx 10^{6}$ GeV, and we take $E_{p,{\rm min}} = 100$ GeV without affecting the results. We find that $e^\pm$ production on photons and synchrotron radiation dominate the cooling of $\pi^\pm$ and $K^\pm$ (see Appendix~\ref{app:timescale}).
The benchmark HE neutrino spectra are shown in Fig.~\ref{fig:nu_flavor}. The competition between $pp$ and $p\gamma$ reactions depends on the effective threshold for meson production and the number density of targets. The $pp$ reaction dominates at low proton energies and produces almost equal neutrino and antineutrino fluxes at $E\lesssim 30$ TeV (left panel) due to equal production of $\pi^+\,(K^+)$ and $\pi^-\,(K^-)$. Once the proton energy exceeds the effective threshold for the $p\gamma$ reaction, this reaction dominates due to the much larger number density of thermal photons and gives rise to higher fluxes of neutrinos than antineutrinos (left panel) due to dominant production of $\pi^+\,(K^+)$ over $\pi^-\,(K^-)$ and stronger synchrotron cooling of $\mu^\pm$. The latter effect also suppresses production of $\nu_e$ and $\bar\nu_e$ from $\mu^\pm$ decay compared to that of $\nu_\mu$ and $\bar\nu_\mu$ from $\pi^\pm\,(K^\pm)$ decay (middle panel). 

To cover a wide range of GRB luminosities,
we compare the all-flavor neutrino spectra in the right panel of Fig.~\ref{fig:nu_flavor} using $L_{\rm iso}=10^{50}$--$10^{53}$~erg/s with other parameters unchanged.
The variations with increasing $L_{\rm iso}$ are caused by the boost of neutrino energy due to a larger Lorentz factor and by more efficient cooling of charged mesons at relatively low energies due to a larger photon density or $B_u$. For $L_{\rm iso}=10^{53}$~ erg/s, cooling by pair production and synchrotron radiation suppresses neutrino production from $\pi^\pm$ decay so much that $K^\pm$ decay produces half of the neutrinos at 100~TeV. 

\section{Annihilation of High-Energy neutrinos with Low-Energy antineutrinos}
Simulations show that the nonrelativistic radioactive outflows occur within a few seconds of jet launching and last for a time comparable to the jet duration   \cite{Just:2022fbf,Fujibayashi:2022xx}. Therefore, most of the HE neutrinos produced at $R_{\rm is} \sim 10^9$--$10^{10}$~cm can meet the LE antineutrinos from the radioactive outflows. The resulting annihilation takes $\lesssim$ 1~s, which is significantly shorter than the outflow duration. We use a one-dimensional Monte Carlo code to simulate the propagation of HE neutrinos in the LE antineutrino background and the accompanying processes, which include $\nu_\alpha\bar\nu_\alpha$ annihilation and decays of the produced $\pi^\pm$ and $K^\pm$. Note that we also include the annihilation of the regenerated HE $\nu_\alpha$ from the earlier annihilation processes. The radial step $\Delta R$ to the next interaction point for an HE $\nu_\alpha$ with energy $E_H$ at radius $R$ is determined by $\tau_\alpha(E_H,R,\Delta R) = -\ln \lambda$, where $\tau_\alpha$ is the optical depth due to $\nu_\alpha \bar\nu_\alpha$ annihilation and $\lambda$ is a random number between 0 and 1. The relevant cross section $\sigma_{\nu_\alpha\bar\nu_\alpha}(s)$ peaks at $s=m_Z^2$, and we use PYTHIA 8.3 \cite{Bierlich:2022pfr} to track the final products for all $s$ values.

Assuming vacuum oscillations of LE antineutrinos prior to annihilation, we show the spectra of HE neutrinos following annihilation in Fig.~\ref{fig:nu_flavor}. The $Z$-resonance starts to occur at $E\sim 300$ TeV. Beyond this energy, the HE neutrino flux drops but the HE antineutrino flux slightly increases due to production from $Z$ decay (left panel). HE neutrinos at these energies are mostly $\nu_\mu$, which are annihilated by the LE $\bar\nu_\mu$ (middle panel). Strikingly, the $\nu_e$ and $\bar\nu_e$ regenerated from such annihilation are even more abundant than those produced without annihilation. Consequently, the flux of $\nu_e+\bar\nu_e$ is very similar to that of the regenerated $\nu_\tau+\bar\nu_\tau$ at $E\gtrsim 300$~TeV following annihilation (middle panel). The above effects produce a unique flavor composition of $(F_{\nu_e+\bar\nu_e}: F_{\nu_\mu+\bar\nu_\mu}: F_{\nu_\tau+\bar\nu_\tau})_\star \approx(1 : 10 : 1)$ for $E\gtrsim 300$~TeV at source, which corresponds to a normalized composition of $\approx(0.21 : 0.42 : 0.37)$ at Earth following vacuum oscillations. The corresponding all-flavor spectrum with the suppression due to annihilation is shown in the right panel of Fig.~\ref{fig:nu_flavor} for different values of $L_{\rm iso}$. With an optical depth of 3--8 for $\nu_\mu\bar\nu_\mu$ annihilation for $E_H\sim 0.4$--1 PeV, the suppression factor can be as high as 3--3.5 (see the inset in the right panel of Fig.~\ref{fig:nu_flavor}). 

\section{Diffuse neutrino flux}
Although prompt HE neutrinos associated with bright GRBs are tightly constrained by IceCube \cite{Abbasi2012_grb,Aartsen2015_grb,Aartsen2016_grb,Aartsen2017_grb,IceCube:2022rlk}, subleading contributions of precursor neutrinos from jet propagation inside stellar matter still remain possible. The most recent analyses showed that precursor neutrinos preceding the prompt $\gamma$-rays by tens of seconds are limited to $\lesssim10$\% of the diffuse flux at $E = 100$ TeV (see Fig.~7 of \cite{IceCube:2022rlk}). On the other hand, LLGRBs, whose rate is $\sim 10$ times higher than that of classical GRBs \cite{Soderberg:2006vh,Liang:2006ci,Nakar:2015tma}, could be caused by a shock breakout driven by a relativistic jet propagating into an extended envelope \cite{Campana:2006xx}. They may have a common collapsar origin and host jets with similar luminosities inside stars to those for bright GRBs \cite{Nakar:2015tma,Senno2015}.
Note that, in the shock breakout scenario, the observed $\gamma$-ray luminosities of LLGRBs ($L_{\gamma, \rm llgrb}^{\rm obs} \sim 10^{46}$--$10^{48}$ erg/s) are much lower than those of bright GRBs because the emission for LLGRBs spans a longer duration ($\Delta T_{\gamma, \rm llgrb} \sim 10^{3}$--$10^{4}$ s) as determined by the large shock breakout radius and is over a wider opening angle ($\theta_{\rm llgrb} \sim 1$ rad). Considering the duration correction, the beam factor correction, and a baryon loading factor $f_p \sim10$, the true luminosity of LLGRB jets initially launched inside stars could be $L_{\rm iso} \sim L_{\gamma, \rm llgrb}^{\rm obs} f_p \frac{\Delta T_{\gamma, \rm llgrb}}{\Delta T_{\rm eng}}\frac{ \theta^2_{\rm llgrb}}{\theta^2_{\rm j}} \sim 10^{51}$--$10^{53}$ erg/s, where $\Delta T_{\rm eng} \sim 10$--100 s is the central engine duration and $\theta_{\rm j} \sim 0.1$ rad is the initial jet opening angle at launch. As HE neutrinos are produced deep inside stars in our work, LLGRBs and bright GRBs are similar sources for HE neutrinos, except that they have different occurrence rates. Further, choked collapsar jets could be even more common than bright GRBs and LLGRBs if the central engine is not active for too long \cite{Sobacchi:2017wcq}.
To incorporate contributions from all similar sources, we introduce the parameter $f_\star$ as the ratio of the total rate of collapsars with mildly-magnetized jets to that of bright GRBs.

\begin{figure}[htbp]
\includegraphics[width=1.0\columnwidth]{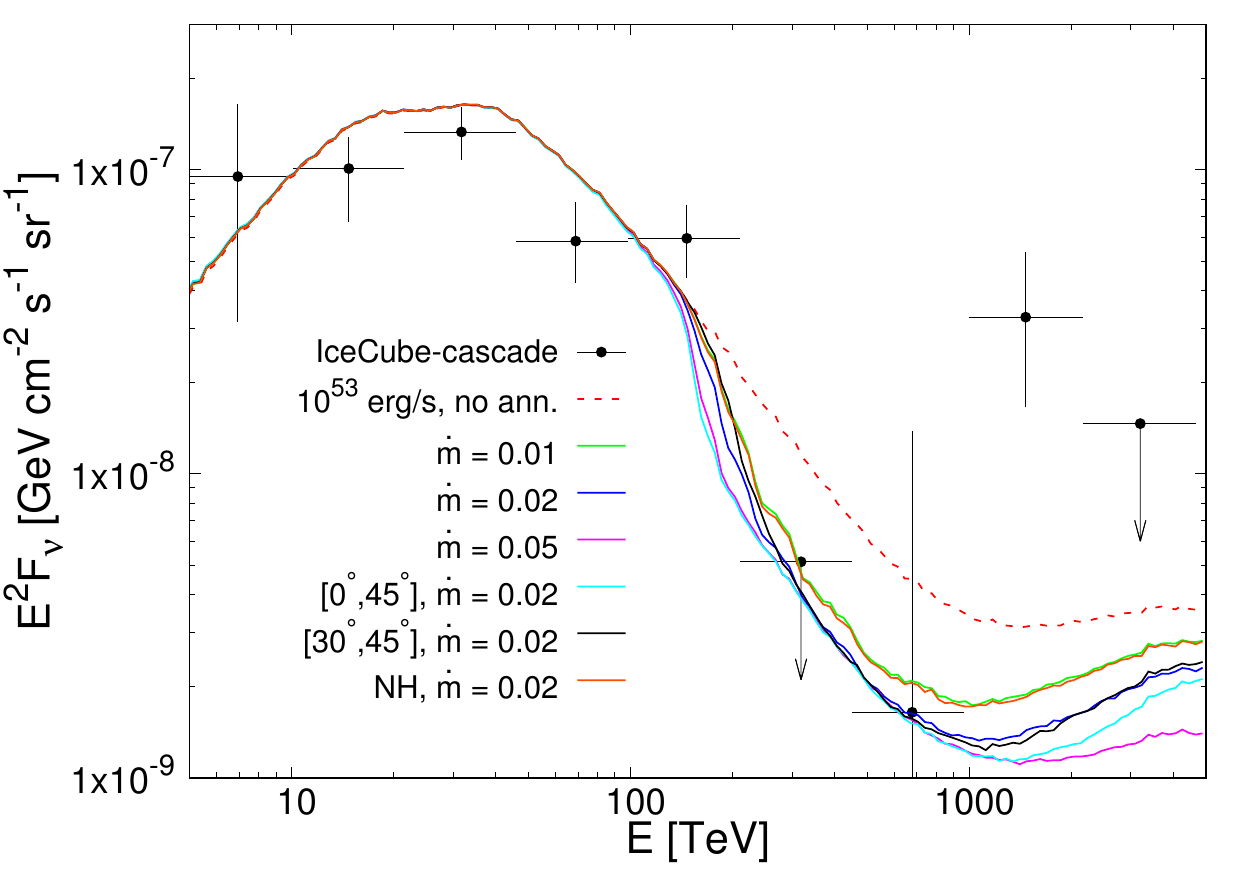}
\caption{All-flavor diffuse neutrino flux spectra from our collapsar model with and without annihilation compared to IceCube data \cite{IceCube:2020acn}. Contributions from jets with $L_{\rm iso,53}=1$ at redshift $z=1$ are assumed to dominate. Results for different values of $\dot m\equiv\dot M/(M_\odot\rm/s)$, asymmetric winds confined to $\theta'=[0^\circ,45^\circ]$ and $[30^\circ,45^\circ]$, and the NH oscillation scenario for LE antineutrinos are also shown.} 
\label{fig:nuflx0}   
\end{figure}

The unnormalized HE neutrino spectrum from a single source is shown in Fig.~\ref{fig:nu_flavor}. To obtain the diffuse neutrino flux from all similar sources over cosmic history, we multiply the unnormalized spectrum by the following factor \cite{Murase2013} 
\begin{align}
&\frac{c}{4\pi H_0}\frac{f_zf_\star\dot{n}_{\rm GRB}f_{\rm cr} E_{\rm iso}}{\ln(E_{p,\rm max}/E_{p,\rm min})}
\frac{1}{{\rm GeV}\ {\rm cm}^{-2}\ {\rm s}^{-1}\ {\rm sr}^{-1}}\nonumber \\
&\sim 7\times 10^{-9}\left(\frac{f_zf_\star\dot{n}_{\rm GRB}}{\rm Gpc^{-3}~yr^{-1}}\right)f_{\rm cr} E_{\rm iso,53},   
\end{align}
where $f_z\sim 3$ accounts for evolution of source population and neutrino energy with redshift $z$ \cite{Waxman:1998yy}, $\dot{n}_{\rm GRB} \sim 1$~Gpc$^{-3}$~yr$^{-1}$ is the rate of bright GRBs per unit volume, $E_{\rm iso}$ is the total isotropic energy of jets, and $f_{\rm cr}$ is the fraction of jet energy converted into HE protons. Taking $f_\star=10$, $f_{\rm cr} E_{\rm iso,53}\sim 1$, and assuming that contributions from sources at $z\sim 1$ dominate, we compare the all-flavor $E^2 F_\nu(E)$ from our model to the IceCube data \cite{IceCube:2020acn} in Fig.~\ref{fig:nuflx0}. It can be seen that LLGRBs and choked jets with collapsar origin may explain the observed TeV--PeV flux at IceCube. More interestingly, the spectrum with suppression at $E\gtrsim 100$--200~TeV due to annihilation (and redshift) leads to better agreement with data than that without annihilation. Results using $\dot M=0.01$ and $0.05~M_\odot$/s are also shown for comparison. Note that spherical winds and vacuum oscillations of LE antineutrinos are assumed for our benchmark study. The annihilation effect could be similar or even enhanced for asymmetric winds while it will be reduced for the NH oscillation scenario with a smaller $P_{\bar\nu_\mu}$ (see Appendices \ref{app:geometry} and \ref{app:oscil}).
We find that annihilation is significant under a range of assumptions (see Fig.~\ref{fig:nuflx0}). Our model, however, cannot account for the PeV events detected at IceCube despite the production of a small bump above $\sim 1$~PeV associated with a minimum in the timescale for cooling of $K^\pm$ by pair production (see Fig.~\ref{fig:Ktimescales} in Appendix~\ref{app:timescale}).
Another source is required for these events \cite{Chen:2014gxa,Palladino:2017qda,Chianese:2017jfa,Denton:2018tdj,Riabtsev:2022ynm}.

\section{Summary and discussion}
Considering collapsars as sources for both HE neutrinos and $r$-process nuclei, we have investigated a novel effect on HE neutrinos caused by their annihilation with LE antineutrinos from the $\beta$-decay of $r$-process nuclei. Assuming typical jet parameters, we have also discussed production of HE neutrinos at internal subshocks induced by mildly-magnetized jets deep inside collapsars. Bright GRBs, LLGRBs, and choked GRBs may have common origin in collapsars, and contributions to the diffuse HE neutrino flux are expected to be dominated by the latter two. Our collapsar model suggests that these sources can well account for the observed flux at IceCube, especially when annihilation with LE antineutrinos from $\beta$-decay of $r$-process nuclei is included. Specifically, an excess at 10--100 TeV and potentially, a deficit at 0.1--1 PeV can be explained consistently for the IceCube events without overproducing the diffuse $\gamma$-ray background (see e.g., \cite{Murase2016,Murase:2019vdl,Capanema:2020rjj,Fang:2022trf}).

A critical test of our HE neutrino production model is detection of precursor HE neutrinos (with a lead time of $\sim 10$--$10^3$ s) from nearby LLGRBs or bright GRBs. For an LLGRB at 100 Mpc with canonical parameters, IceCube-Gen2 can detect $\sim 10$ cascade events within 10--100 s if the jets are mildly magnetized. This short emission would also favor the speculation that LLGRBs host very similar jets to those in bright GRBs. Our model predicts a flavor composition at Earth close to that in the so-called muon-damped scenario at 10~TeV to 1~PeV, which can be distinguished from those in the standard pion and neutron decay scenarios \cite{Ackermann:2017pja,IceCube-Gen2:2020qha,Song:2020nfh} by precise measurements. The expected spectrum decays as $E^{-\gamma}$ at energies below $\sim 100$~TeV with $\gamma\sim 0.5$--1, followed by a steepening above $100$--200 TeV due to the annihilation effect. Anticipating 10 years of data at IceCube-Gen2, we find that a spectral steepening with $\Delta\gamma>1$ for our scenario can be observed at 95\% CL (see Appendix~\ref{app:gen2}).
In contrast, no clear break is expected without annihilation. The above neutrino signatures together with the $r$-process imprints in collapsar lightcurves can provide important support of collapsars as sources for both HE neutrinos and $r$-process nuclei.
 
We have made several simplifications in this initial study. Future improvements include detailed modeling of LE antineutrino emission, spatial and temporal variations of $r$-process production in collapsars, and self-consistent collapsar simulations for HE neutrino production. Finally, the same annihilation effect should occur in BNSMs that host short GRBs and the $r$-process, and can in principle be tested with a future nearby event.

\begin{acknowledgments}
We thank 
Brian Metzger and  
Daniel Siegel for useful discussions. We are also grateful to the anonymous referees for their constructive comments. This work was supported in part by the National Natural Science Foundation of China (12205258) and the Natural Science Foundation of Shandong Province, China [ZR2022JQ04 (G.G.)], the US Department of Energy [DE-FG02-87ER40328 (Y.Z.Q.)], and the National Science and Technology Council (No.~110-2112-M-001-050, 111-2628-M-001-003-MY4), the Academia Sinica (No.~AS-CDA-109-M11), and the Physics Division of the National Center for Theoretical Sciences, Taiwan (M.R.W.).
\end{acknowledgments}




\appendix
\section{Geometry of disk winds}\label{app:geometry}

We have assumed in the main text that the disk winds are spherically symmetric. These winds could be non-spherical and be channelled into a certain range of angles off the jet axis. For example, Ref.~\cite{MacFadyen:1999xx} found that $\theta'$ (see  Fig.~\ref{fig:sketch} of the main text) could range from $\sim 10^\circ$ to $\sim 40^\circ$, while Ref.~\cite{Siegel:2018zxq} assumed $30^\circ \le \theta' \le 45^\circ$. A collimated wind was also considered in Ref.~\cite{Hayakawa:2018uxm} 
with $\theta'\le 30^\circ$. Approximating asymmetric winds as ``isotropic'' winds confined to $\theta'=[\theta'_{\rm min},\theta'_{\rm max}]$, we have shown the results in Fig.~\ref{fig:nuflx0} of the main text. For winds confined within $[0^\circ, 45^\circ]$, the LE neutrino intensity is enhanced by a factor of $3$ compared to the spherical case; while for $\theta'=[30^\circ,45^\circ]$, the intensity and the annihilation effect are close to the spherical case.

\section{Oscillation scenarios for LE antineutrinos}\label{app:oscil}

Oscillations of LE antineutrinos prior to interaction with HE neutrinos depend on the local densities at the $\bar\nu_e$ emission site and the interaction site.
We consider annihilation inside the jets where the typical density is far below 1~g/cm$^3$. The wind density at the $\bar\nu_e$ emission site is $\rho \sim \dot M/(4\pi r^2v_{\rm ej}) \sim 10^3$~g/cm$^3$ at $r\sim 10^9$~cm assuming $\dot M \sim 0.01 M_\odot$/s and $v_{\rm ej}\sim 0.05c$. On the other hand, the density at the emission site could be dominated by stellar matter and can reach $\sim 10^6$~g/cm$^3$
at $10^9$ cm, although stellar matter may be blown away by the jets and/or disk winds. The relevant density could also be low if $\bar\nu_e$ are emitted at larger radii. Given these complexities and uncertainties, we consider the extreme cases where $\bar\nu_e$ are emitted at very high and very low densities, respectively.  

The $\bar\nu_e$, if emitted at densities of $\gg 10^4$~g/cm$^3$, almost coincides with the effective mass eigenstate in matter $\bar\nu_{1m}$ or $\bar\nu_{3m}$ for normal (NH) or inverted (IH) neutrino mass hierarchy, respectively. For adiabatic evolution during propagation, the emitted $\bar\nu_e$ becomes the vacuum mass eigenstate $\bar\nu_{1}$ (NH) or $\bar\nu_{3}$ (IH) at very low densities for the interaction site. The probability $P_{\bar\nu_\alpha}$ to be a $\bar\nu_\alpha$ at the interaction site is $(P_{\bar\nu_e}, P_{\bar\nu_\mu}, P_{\bar\nu_\tau})\approx (0.675, 0.095, 0.23)$ (NH) or (0.022, 0.545, 0.433) (IH) with the mixing parameters from Ref.~\cite{PDG:2022pth}.
If $\bar\nu_e$ are emitted at very low densities, $(P_{\bar\nu_e}, P_{\bar\nu_\mu}, P_{\bar\nu_\tau})\approx(0.55, 0.18, 0.27)$ as expected from pure vacuum oscillations. As HE neutrinos are mainly $\nu_\mu$ in our model, the flux of LE $\bar\nu_\mu$ is the most relevant for annihilation. The NH and IH scenarios are the most pessimistic and optimistic, respectively. For $\bar\nu_e$ emitted at intermediate densities, $P_{\bar\nu_\mu}$ ranges from 0.095 to 0.545. Compared to pure vacuum oscillations, the NH scenario reduces the LE $\bar\nu_\mu$ flux by a factor of $\sim 2$. As shown in Fig.~\ref{fig:nuflx0} of the main text, the NH scenario with $\dot m=0.02$ gives rise to a similar annihilation effect to that for pure vacuum oscillations with $\dot m=0.01$.

\section{Timescales for acceleration and cooling}\label{app:timescale}

\begin{figure}[htbp]
\includegraphics[width=1.0\columnwidth]{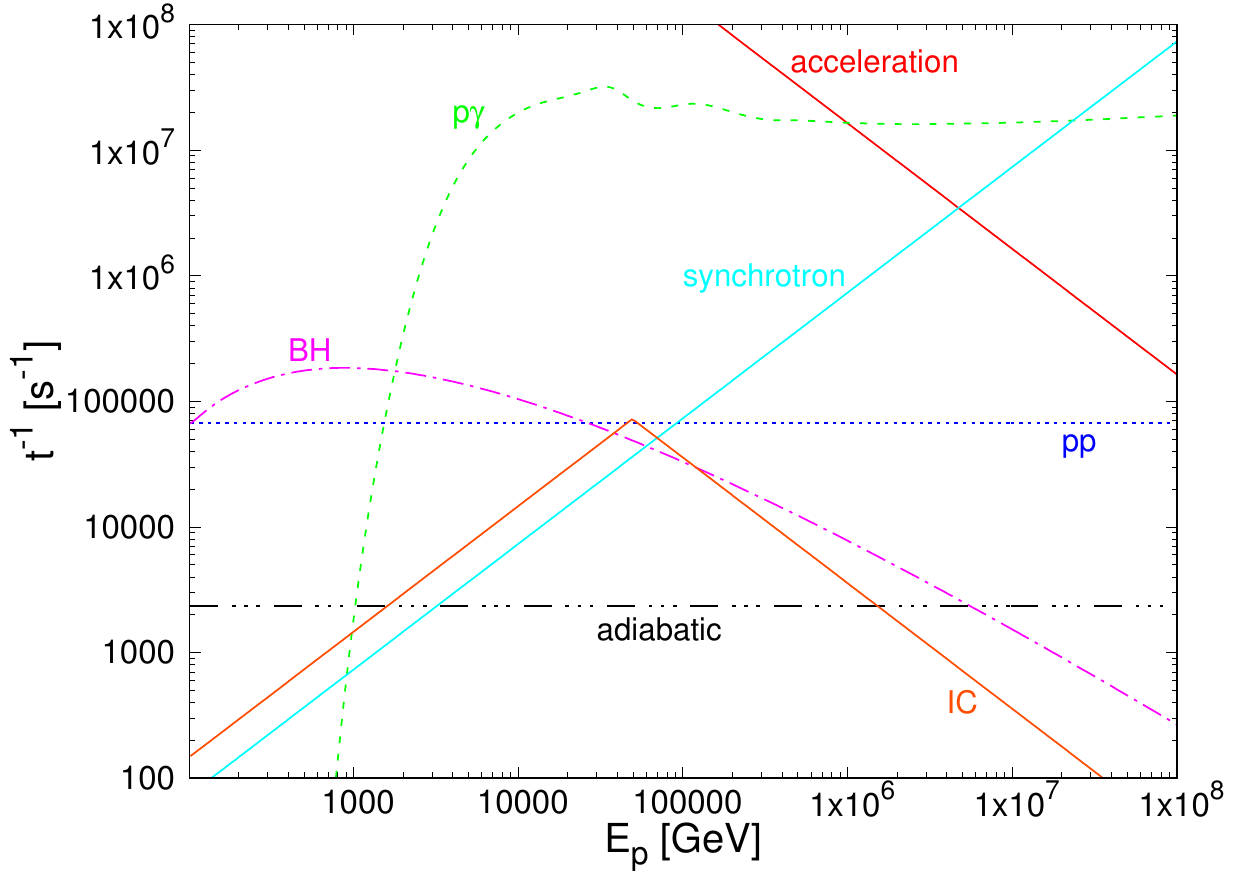}
\caption{Timescales for shock acceleration of protons and their cooling by various processes in the rest frame of the shocked jet. We have taken $L_{\rm iso}=10^{53}$ erg/s and the benchmark values for the other parameters.}
\label{fig:ptimescales} 
\end{figure}

Figure~\ref{fig:ptimescales} shows the timescales for proton acceleration and cooling in the rest frame of the shocked jet.
The acceleration timescale is approximated by $E_p/(e B_d c)$. The cooling timescale due to $p\gamma$ reactions is given by 
\begin{align}
t^{-1}_{p\gamma}(E_p)
= &c\int d\epsilon d\Big({\cos\theta_{p\gamma} \over 2}\Big) \nonumber \\ 
& \times \kappa(\tilde {\epsilon})\sigma_{p\gamma}(\tilde{\epsilon}) {dn(\epsilon) \over d\epsilon}(1-\cos\theta_{p\gamma}),  \label{eq:tpg}
\end{align}
where $\theta_{p\gamma}$ is the intersection angle, $dn(\epsilon)/d\epsilon$ is the differential density of thermal photons,
$\tilde{\epsilon}$ is the photon energy in the rest frame of the proton, and $\kappa$ and $\sigma_{p\gamma}$ are the inelasticity and cross section, for which we take the expressions
from \cite{Guo:2019ljp}. The cooling timescale due to the Bethe-Heitler (BH) process can be obtained similarly and we use the inelasticity and cross section from \cite{Chodorowski:1992}.
For the cooling timescales due to $pp$ scattering, synchrotron radiation, inverse Compton (IC) scattering, and adiabatic expansion, we use the expressions given in Ref.~\cite{Razzaque2005}. For the parameters chosen, the maximal proton energy achieved in the comoving frame of the shocked jet is about $10^6$ GeV. 

\begin{figure}[htbp]
\includegraphics[width=1.0\columnwidth]{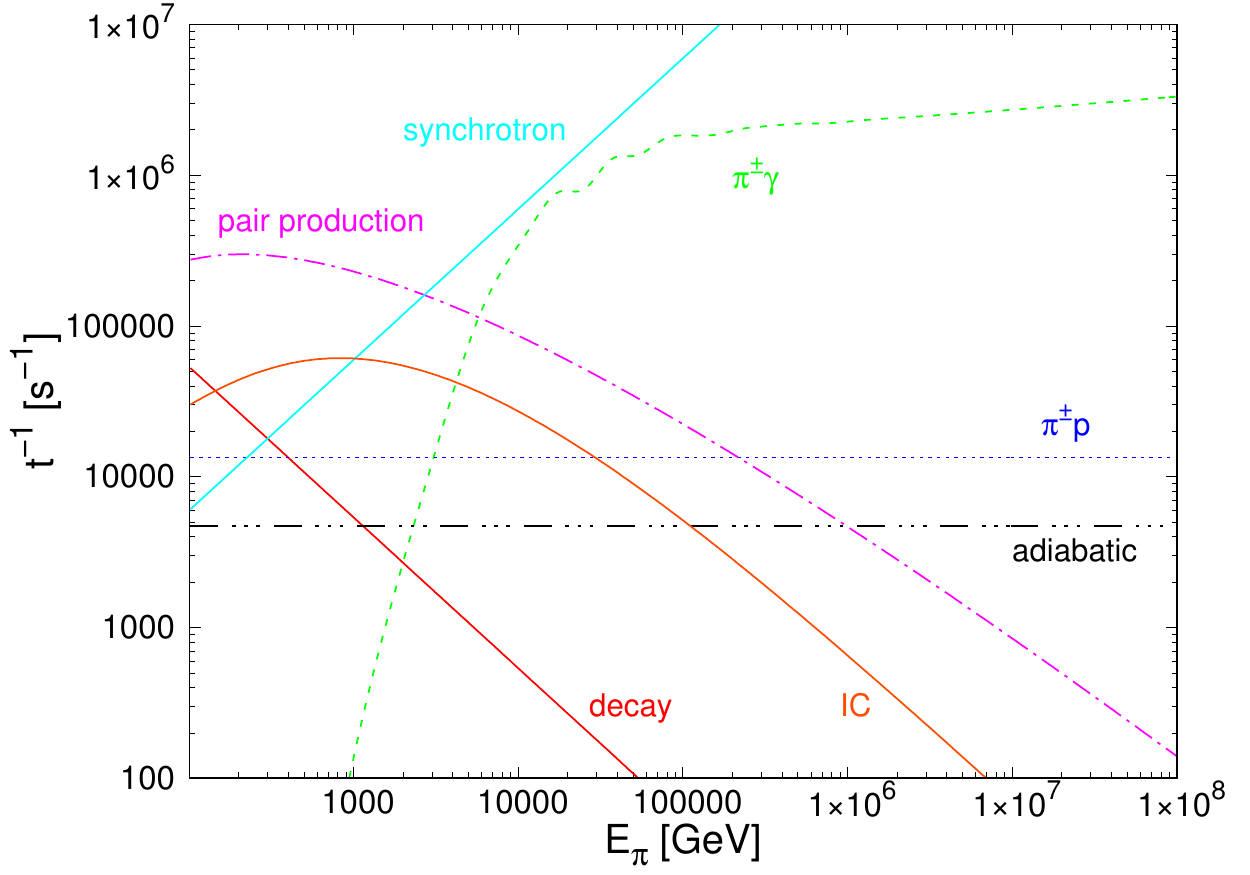}
\caption{Timescales for $\pi^\pm$ decay and their cooling by various processes in the rest frame of the unshocked rapid jet. We have taken $L_{\rm iso}=10^{53}$ erg/s and the benchmark values for the other parameters.}
\label{fig:pitimescales}    
\end{figure}

\begin{figure}[htbp]
\includegraphics[width=1.0\columnwidth]{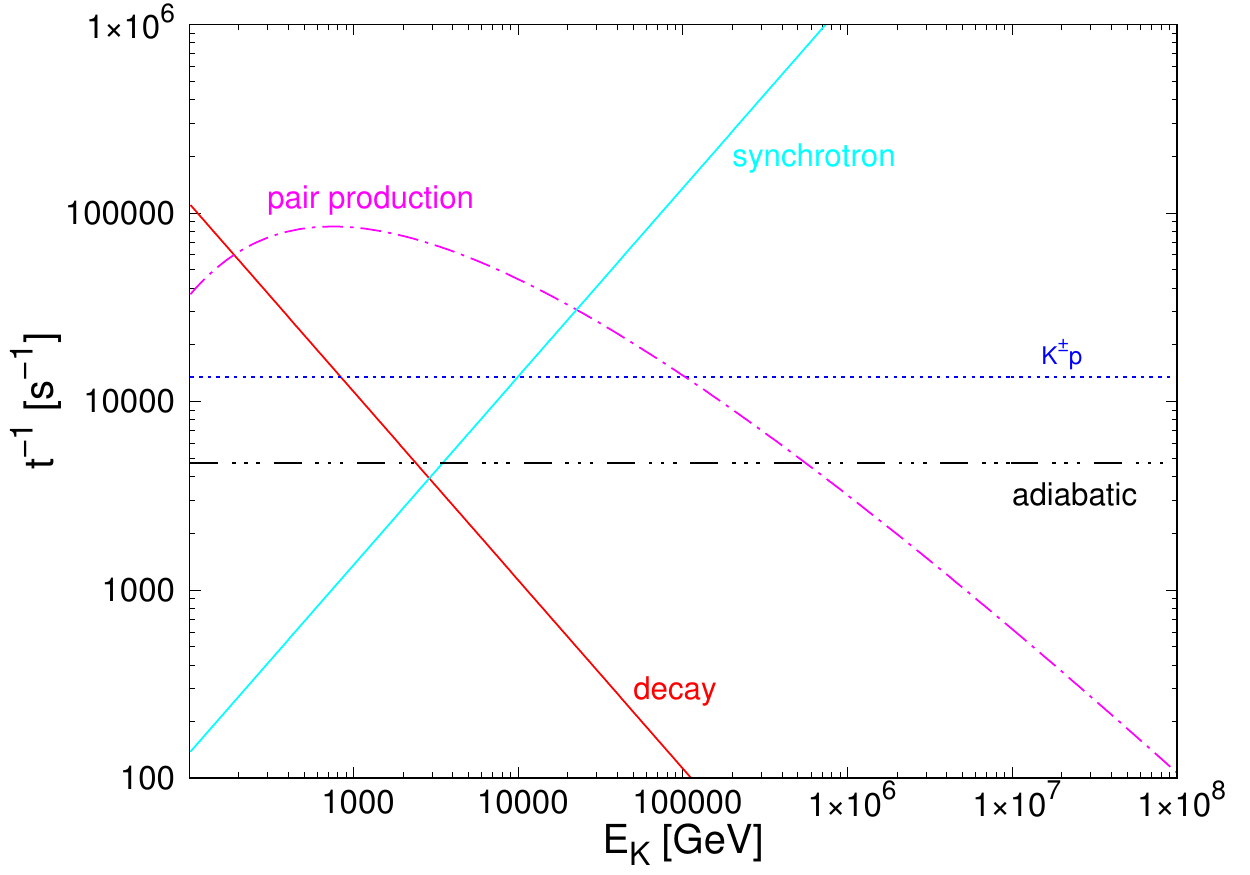}
\caption{Timescales for $K^\pm$ decay and their cooling by various processes in the rest frame of the unshocked rapid jet. We have taken $L_{\rm iso}=10^{53}$ erg/s and the benchmark values for the other parameters.}
\label{fig:Ktimescales}    
\end{figure}

To avoid strong synchrotron cooling of mesons, we consider $\pi^\pm$ and $K^\pm$ production via $pp$ and $p\gamma$ reactions in the unshocked rapid jet. Figure~\ref{fig:pitimescales} shows the timescales for $\pi^\pm$ cooling by various processes. Synchrotron radiation, $e^\pm$ pair production on photons, $\pi^\pm p$ collision, and adiabatic cooling are calculated similarly to the case of protons. For $\pi^\pm\gamma$ process, we use the cross sections for IC scattering and multi-meson production (dominated by $\gamma + \pi^\pm \to \pi^0+\pi^\pm$ at low energy) from \cite{Kaiser:2008ss}. Pair production is studied similarly to that for protons using the cross section and inelasticity from \cite{Chodorowski:1992}. Note that the expressions in \cite{Chodorowski:1992} apply equally to proton and charged mesons, as they were derived for pair production in a Coulomb field \cite{Maximon68}. As the observed neutrino energy $E_{\rm ob} \sim \Gamma_r E_\pi/4$, both pair production and synchrotron radiation are relevant for determining the HE neutrino fluxes at $0.1 \lesssim E \lesssim 1$ PeV. Cooling timescales for $K^\pm$ have been estimated similarly to those for $\pi^\pm$ (see Fig.~\ref{fig:Ktimescales}) except that $K^\pm \gamma$ multi-meson production and IC are ignored compared to pair production and synchrotron radiation.      

\section{Detecting spectral steepening with IceCube-Gen2}\label{app:gen2}

With a much better energy resolution, the cascade events at IceCube-Gen2 can be used to explore the potential of observing a spectral break in the HE neutrino flux \cite{IceCube:2021jmr,Kochocki:2023lhh}. Lacking the detailed detector response of IceCube-Gen2, we carry out a relatively crude study. The only information we rely on is the cascade effective area of IceCube-Gen2 from \cite{IceCube-Gen2:2020qha}, which is about 1 order of magnitude larger than that of IceCube. Note that for the cascade effective area given, the astrophysical HE neutrinos are typically assumed to be distributed isotropically in sky and equally among all flavors. Processes producing cascade events include neutral-current (NC) reactions of all flavors, charged-current (CC) reactions of $\nu_e+\bar\nu_e$, and part (with a probability of $\sim 83$\%) of the CC reactions of $\nu_\tau+\bar\nu_\tau$. The effective areas for the CC and NC reactions of HE neutrinos at a given zenith angle $\theta_z$ can be obtained with consideration of Earth absorption. The deposited energy $E_d$ in the detector depends on the neutrino energy $E$ and the interaction type, and we follow the expressions [Eqs.~(B16-B23)] given in Ref.~\cite{Palomares-Ruiz:2015mka} to calculate $E_d$. When generating the $E_d$ distribution, we also consider an energy resolution of 10\%. The expected spectrum of $E_d$ can then be obtained for any given astrophysical neutrino flux.

As the atmospheric muon background will be strongly suppressed by veto \cite{Ackermann:2017pja,IceCube:2020acn,IceCube-Gen2:2020qha,Kochocki:2023lhh}, we only consider the background from atmospheric neutrinos, using the fluxes from the MCEq tool \cite{Fedynitch:2015zma}. The expected $E_d$ spectrum for this background is calculated similarly to that for astrophysical neutrinos. 
The veto can also be used to suppress the atmospheric neutrino background \cite{Ackermann:2017pja,IceCube-Gen2:2020qha,Kochocki:2023lhh,IceCube:2020acn}, and we take an overall suppression factor of $1/3$ for $\cos\theta_z\ge 0.2$ (see e.g., Fig.~4.10 of \cite{Niederhausen:2018}).

\begin{figure}[htbp]
\includegraphics[width=1.0\columnwidth]{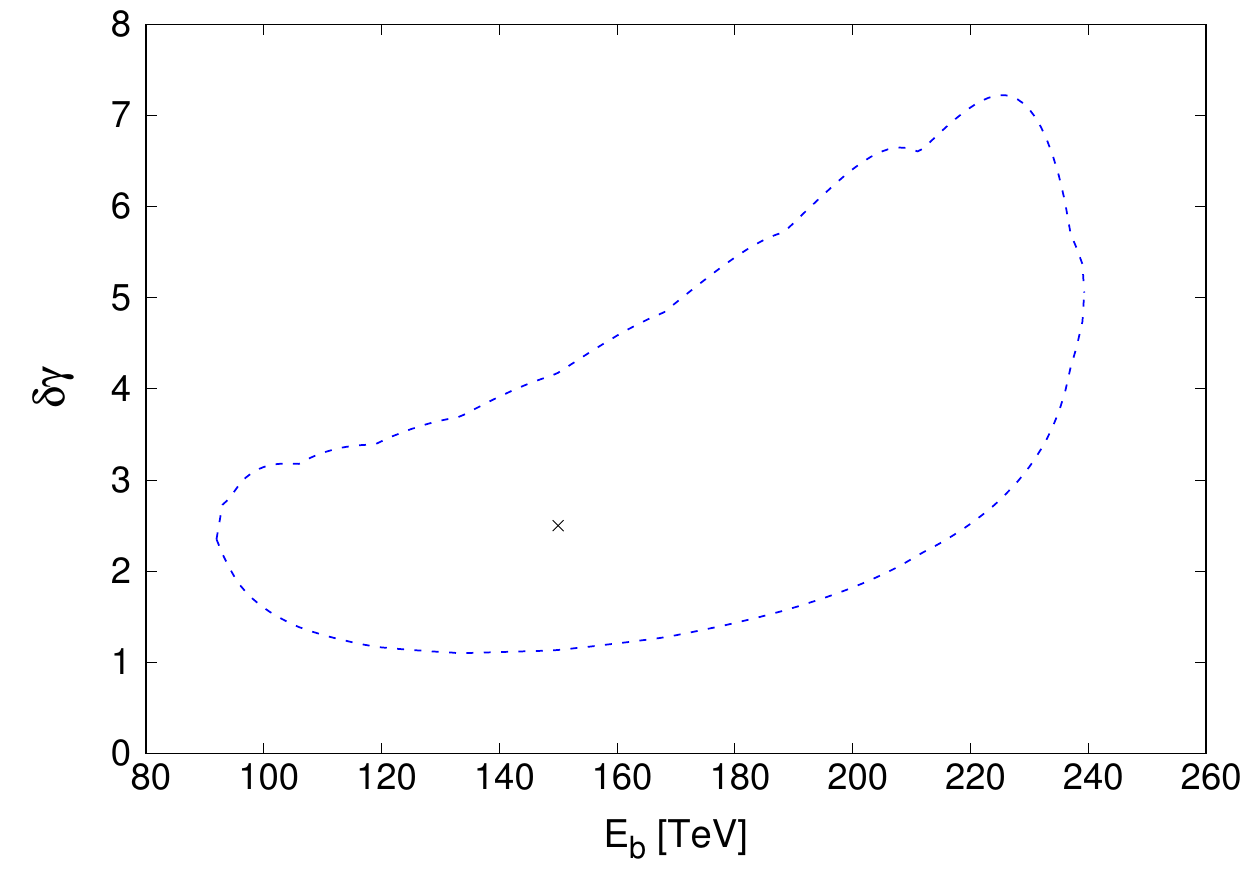}
\caption{Sensitivity for measuring the break energy $E_b$ and the spectral index change $\delta\gamma$ at 95\% CL with 10 years of IceCube-Gen2 data. The $\times$ symbol represents the best-fit point with $\chi^2(E_b, \delta\gamma)=\chi^2_{\rm min}$.}
\label{fig:spec_sen} 
\end{figure}

Based on Fig.~\ref{fig:nuflx0} of the main text, we simply model our astrophysical HE neutrino flux with a broken power-law at $10~{\rm TeV}\le E \le 500$ TeV:
\begin{equation}
E^2 F_\nu = A \times 
\begin{cases} 
({E/100~\rm TeV})^{-\gamma}, & E < E_b, \\
({E_b/ 100~\rm TeV})^{-\gamma}({E/E_b})^{-\gamma-\delta \gamma}, & E > E_b. 
\end{cases} 
\end{equation}
We generate the so-called Asimov dataset assuming $A_0=2\times 10^{-8}~\rm GeV~cm^{-2}~s^{-2}~sr^{-1}$, $E_{b0}=150$ TeV, $\gamma_0=0.5$, and $\delta\gamma_0=2.5$. 
We take 30
equal-size bins for $E_d$ between 10 TeV and 500 TeV on logarithmic scale, and 10 equal-size bins for $\cos\theta_z=[-1, 1]$.
The $\chi^2$ function is defined as 
\begin{align}
\chi^2(E_{b}, \delta\gamma) = \min_{\{A, \gamma,\alpha\}} \Big[ -2\sum_{i,j} (n_{ij} \ln{\mu_{ij}}-\mu_{ij})\Big],   \label{eq:chisq}
\end{align}
where $n_{ij}$ is the event number of the Asimov dataset in the $i$th $E_d$-bin and the $j$th $\cos\theta$-bin, and $\mu_{ij}(A, E_b, \gamma, \delta\gamma, \alpha)$ is the expected event number with $\alpha$ being the relative uncertainty of the atmospheric neutrino background. Note that $n_{ij}=\mu_{ij}(A_0, E_{b0}, \gamma_0, \delta\gamma_0, 0)$. As suggested by Eq.~(\ref{eq:chisq}), the parameters $A$, $\gamma$, and $\alpha$ are allowed to vary freely to minimize the $\chi^2$ function for a given combination of $E_b$ and $\delta\gamma$.   

The $E_b$-$\delta\gamma$ contour in Fig.~\ref{fig:spec_sen} shows the sensitivity for measuring the spectral break with 10 years of IceCube-Gen2 data.
This contour corresponds to $\chi^2(E_b, \delta\gamma)-\chi^2_{\rm min} \approx 5.99$ (95\% CL). We see that a steepening of the spectrum with $\delta \gamma>1$ at $100~{\rm TeV}\lesssim E_b\lesssim 240$~TeV can be measured.

%

\end{document}